\documentclass[aps,floats,epsf,twocolumn]{revtex4}
 \usepackage[T1]{fontenc}
\usepackage[latin1]{inputenc}
\usepackage{graphics}
\usepackage{amssymb}
\usepackage{amsmath}
\usepackage{graphicx,epsf}
\begin{document}

\title{Coulomb interaction at the metal-insulator critical point in graphene}

\author{Vladimir  Juri\v ci\' c$^1$, Igor F. Herbut$^{1,2}$,  and Gordon W. Semenoff$^3$}

\affiliation{$^1$ Department of Physics, Simon Fraser University,
 Burnaby, British Columbia, Canada V5A 1S6 \\ $^2$ Kavli Institute for
 Theoretical Physics, University of California, Santa Barbara, CA
 93106, USA \\ $^3$ Department of Physics and Astronomy, University of
 British Columbia, 6224 Agricultural Road, Vancouver, British
 Columbia, Canada V6T 1Z1}

\begin{abstract} We compute the renormalization group flow of the
long-ranged electron-electron interaction at the Gross-Neveu quantum critical point between the semimetal and
the excitonic insulator in graphene, perturbatively in the small parameter $\epsilon=d-1$, with $d$ as the
spatial dimension. The $O(\epsilon)$ correction to the usual beta-function makes the long-range interaction
only more irrelevant at the critical than at the Gaussian fixed point. A weak long-range tail of the Coulomb
interaction is found to be marginally irrelevant also in arbitrary dimension when the number of Dirac fermions
is large. Its ultimate irrelevancy notwithstanding, it is shown that the metal-insulator transition may still
be induced by increasing only the long-range tail of the Coulomb interaction.

\end{abstract}
\maketitle

\vspace{10pt}

   Increasing the strength of electron-electron interactions relative
   to the bandwidth is expected to transform graphene from its usual
   semi-metallic phase into the gapped Mott insulator.  This quantum
   phase transition has been studied by a variety of numerical
   \cite{sorella, martelo, paiva, drut} and analytical \cite{gorbar,
   leal, herbut1, herb-jur-roy, herb-jur-vaf} techniques, and it
   represents a condensed matter analog of the particle physics
   phenomenon of chiral symmetry breaking. Graphene provides
   2+1-dimensional Dirac fermions \cite{Semenoff:1984dq} where the
   presence of two valleys and two spin states result in $SU(4)$
   chiral symmetry, which can be broken to $SU(2)\times SU(2)$ while
   still preserving parity and time reversal invariance.  Similar
   systems have been studied
   \cite{Pisarski:1984dj}-\cite{Semenoff:1990hr} as toy models of
   strong coupling behavior in quantum chromodynamics and technicolor
   theories.  In graphene, this results in breaking of the sublattice
   symmetry and gapping the electron spectrum. Generation and control
   of such a gap is of great importance to potential applications in
   electronics \cite{mindthegap}.

   One issue of contention is the precise role in the mechanism of the
   transition of the long-ranged $\sim 1/r$ tail of Coulomb interaction,
   which remains unscreened when graphene is at the Dirac point.
   Whereas the initial analytical calculations based on the
   Schwinger-Dyson equations \cite{gorbar, leal} found that the
   long-range interaction is crucial and leads to an essential
   singularity in the free energy, recent numerical calculations
   \cite{drut} find only a regular second-order transition.  The
   analytic expansion around the exactly solvable three-dimensional
   limit of the theory \cite{herb-jur-vaf,Semenoff:1990hr} also yields
   a regular critical point at which a weak long-range interaction may
   be shown to represent a marginally irrelevant perturbation.  In the
   present study we complement these results by showing that a weak
   unscreened $\sim 1/r$ tail of the electron-electron repulsion
   remains a marginally irrelevant perturbation for an arbitrary
   number of Dirac fermion components $N$ at the metal-insulator
   quantum critical point near one spatial dimension, as well as for
   large $N$ in arbitrary spatial dimension. This is accomplished by
   computing the {\it first correction} to the beta-function of the
   long-range coupling constant, which we shall henceforth call the
   {\it charge}.  This correction is $O(\epsilon)$ relative to the
   leading term, with $\epsilon=d-1$, and its sign is such that the
   flow of the charge towards zero, although logarithmic in either
   case, becomes {\it faster} at the quantum critical than at the
   Gaussian fixed point (Fig. 1). The renormalization-group flow also
   implies that the transition may be tuned by the charge alone, and
   can then occur even at a sub-critical value of the short-range
   interaction.

   \begin{figure}[t]
{\centering\resizebox*{75mm}{!}{\includegraphics{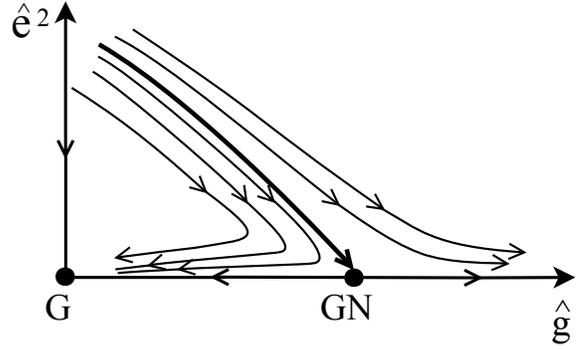}}
\par} \caption[] {Schematic renormalization-group flow in the infrared for the
metal-insulator transition in graphene. $\hat{g}$ is the (dimensionless) short-range, and $\hat{e}^2$ the
long-range component of the Coulomb electron-electron interaction. GN and G are the critical Gross-Neveu and
the Gausian fixed points, respectively. }
\end{figure}

  We consider the simplest Gross-Neveu Lagrangian that should suffice
   to describe the quantum phase transition into the gapped excitonic
   (charge-density-wave) phase with increase of nearest-neighbor
   repulsion in graphene, tuned to be at the Dirac point
   \cite{herbut1, herb-jur-roy},
  \begin{eqnarray}
  L= \bar{\Psi}_\alpha \gamma_0 (\partial_0 + i a)\Psi_\alpha + v
  \bar{\Psi}_\alpha \gamma_i \partial_i \Psi_\alpha \\ \nonumber - g (
  \bar{\Psi}_\alpha \Psi_\alpha ) ^2 + \frac{1}{2 e_d ^2} a
  |\nabla|^{d-1} a.
  \end{eqnarray}
  The short-range coupling $g>0$ is proportional to the
  nearest-neighbor repulsive interaction \cite{herbut1}, and the
  four-component Dirac fermion $\Psi_\alpha $ is defined using the
  conventions of \cite{herbut1, herb-jur-roy}.  For generality we will
  assume an arbitrary number of Dirac fermions $N$, so $\alpha
  =1,2,...N$.  For physical spin-1/2 electrons $N=2$.  The Dirac
  matrices satisfy the standard Clifford algebra in Euclidian
  space-time $\{ \gamma_\mu, \gamma_\nu \} = 2 \delta_{\mu\nu}$. For
  simplicity, we do not consider the transition to a spin-density wave
  phase \cite{herbut1, herb-jur-vaf}.  Our calculation can easily be
  modified to consider that case as well and the conclusion would be similar.

Integration over the gauge field $a$ introduces the instantaneous density-density long-range interaction
\begin{equation}
\bar{\Psi}_\alpha \gamma_0 \Psi_\alpha (\vec{x},\tau) \frac{e^2}{2|\vec{x}-\vec{y}| } \bar{\Psi}_\alpha
\gamma_0 \Psi_\alpha (\vec{y},\tau),
\end{equation}
in any spatial dimension $d$, provided that we define
\begin{equation}
e_d ^2 = e^2 (4\pi)^{\epsilon/2} \Gamma(\epsilon/2),
\end{equation}
and $\epsilon=d-1$. Note that as $\epsilon\rightarrow 0$ the coupling $e_d$ defined this way diverges. This
reflects the fact that the Fourier transform of $\sim 1/r$ interaction in $d=1$ is $\sim \ln(k)$ \cite{bose}.
We will confine out attention to the region $d>1$ in the following.

Let us first outline our method and principal result. Since the inverse of the gauge-field propagator is a
non-analytic function of the momenta for dimensions $1<d<2$,  the coupling  $e_d$ cannot renormalize
\cite{herbut2}, and
  \begin{equation}
  \frac{d e ^2}{d\ln \Lambda} = 0,
  \end{equation}
  where $\Lambda $ is the ultraviolet cutoff.  The instantaneous
long-range interaction does not respect the Lorentz invariance, however, and consequently renormalizes the
Fermi velocity. At a momentum scale $k \ll \Lambda$ one expects the velocity to become
\begin{equation}
v(k) = v + a_d e_d ^2 \ln( \frac{\Lambda}{k}) + b_d e_d ^2 \hat{g} [\ln (\frac{\Lambda}{k})]^ n + O(\frac{e_d
^4}{v}, e_d ^2 \hat{g}^2),
\end{equation}
where $a_d$ and $b_d$ are $d$-dependent numerical coefficients, and $\hat{g} = g \Lambda^\epsilon / \pi$ is the
{\it dimensionless} short-range coupling.  The coefficient $a_d$ for $d=2$ has been computed earlier
\cite{herbut1, vafek, gonzalez, hjv}. Since $v(k)$ is a physical observable it must be independent of the
arbitrary cutoff $\Lambda$, i. e.
\begin{equation}
\frac{d v(k)} {d\Lambda} = 0,
\end{equation}
at {\it all} momenta $k$. This condition of renormalizability of the field theory in Eq. (1) can be satisfied
only if the power of the logarithm in the third term in Eq. (5) is $n=1$, and if $d\hat{g}/d\Lambda=0$. Since
the beta-function for the Gross-Neveu coupling, as it will be shown later, reads
\begin{equation}\label{GN-beta}
\frac{d\hat{g}}{d\ln\Lambda}= \epsilon\hat{g}+(1-\frac{2}{\epsilon})\hat{g} \hat{e}^2 - 2 (2 N-1) \hat{g}^2 + O
(\hat{g}^2{\hat e}^2,\hat{g}^3),
\end{equation}
where $\hat{e}^2=e^2/\pi v$ is the {\it dimensionless charge} which measures the strength of the long-range
coupling relative to the Fermi velocity, the condition of renormalizability at $\hat{e}^2=0$ is satisfied only
at the fixed points: 1) Gaussian, $\hat{g}_G =0$, and 2) the critical $\hat{g}_c= \epsilon/ (2 (2N-1))
+O(\epsilon^2)$. Eqs. (5) and (6) yield then the beta-function for the Fermi velocity
\begin{equation}
\frac{d v}{d\ln \Lambda}= -e_d ^2( a_d +  b_d  \hat{g}^*).
\end{equation}
$\hat{g}^* $ is one of the above two fixed points. We can then recast Eqs. (4) and (8) together as the equation
for the dimensionless charge
\begin{equation}
\frac{d\hat{e}^2}{d\ln \Lambda}= \hat{e}^4 \pi (4\pi)^{\epsilon/2} \Gamma( \epsilon/2 ) ( a_d + b_d \hat{g}^*).
\end{equation}
Both of the coefficients $a_d$ and $b_d$ turn out to be $O(\epsilon)$. Therefore the divergent $\sim
1/\epsilon$ prefactor in the last equation cancels. We thus find a regular expansion of the beta-function for
the dimensionless charge in powers of $\epsilon$ and $\hat{e}^2$,
\begin{equation}\label{Coulomb-beta}
\frac{d\hat{e}^2}{d\ln \Lambda} = \hat{e}^4 \left( 1 + \frac{ \epsilon}{ 2 (2N-1) } + O(\epsilon^2 /N) \right)
+ O (\hat{e}^6 ).
\end{equation}

There are two notable features of the last expression. First, the $O(\epsilon)$ correction has the same sign as
the leading term. As a result, the long-range interaction is in fact more irrelevant in the infrared, although
still marginally so, than  at the Gaussian fixed point. Second, the correction is $O(1/N)$. Therefore, for a
large number of Dirac fermions, the Coulomb interaction is also marginally irrelevant in all dimensions.
Together with the previous analysis near three spatial dimensions \cite{herb-jur-vaf} this means that a weak
long-range tail of Coulomb interaction is an irrelevant perturbation at the Gross-Neveu metal-insulator quantum
critical point in {\it all} perturbatively accessible regimes of the theory.

The beta-function for the short-range coupling in Eq. (7), besides the usual terms \cite{herbut1,herb-jur-roy}
obtains the contribution
 from the long-range Coulomb interaction. The sign of this term
is {\it negative}, so that there is a trajectory in the region ${\hat g}<{\hat g}_c$ that flows right into the
critical point and separates the flows towards the semimetallic and the insulating ground states, as depicted
in Fig.\ 1. The negative sign is crucial to this result, which we may have expected on physical grounds: that
an increase of the charge alone should take the semimetal towards the insulator.  This situation may be
analogous to the one in $3+1$-dimensional quantum electrodynamics, where the chiral-symmetry-breaking
transition may also be tuned by the electromagnetic charge, but the critical behavior seems, on the other hand,
to be controlled by the Nambu-Jona-Lasinio theory with only short-range interactions
\cite{gockeler,kogut-strouthos}.

Let us now present the calculational details. The logarithmic terms in Eq. (5) derive from the diagrams in
Figs.\ (2)-(3). Diagrams in Fig. (3), however, do not contribute to the renormalization of the Fermi velocity.
Since the fixed-point interaction is a contact interaction, the diagram in Fig. (3b) vanishes, whereas the
diagram in Fig. (3c) is independent of the external momentum. The diagram in Fig. (3a), which is also the only
diagram to this order which would be of order $N$, vanishes as well because it is proportional to the factor $
Tr \gamma_0 =0$ that accompanies the fermion loop. This implies that both the coefficients $a_d$ and $b_d$ as
far as the number of Dirac fermions $N$ is concerned are $O(1)$.  Since the fixed point interaction is
$\hat{g}_c \sim 1/N$, it guarantees that the first correction to the Gaussian result is also $O(1/N)$.

   \begin{figure}[t]
{\centering\resizebox*{60mm}{!}{\includegraphics{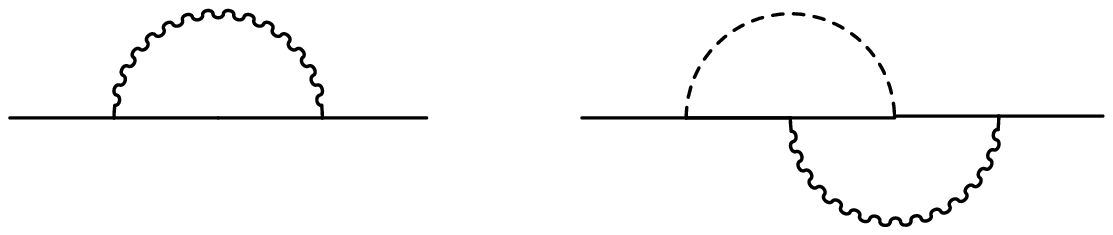}}
\par} \caption[] {One- and two-loop diagrams that renormalize the
Fermi velocity. The wavy line is the long-range interaction, and the dashed line is the contact fixed-point
interaction.}
\end{figure}

   One is therefore left with the diagrams in Fig. (2) to
   compute. Defining the self-energy from the Dirac fermion propagator
   $G(k)$ as
   \begin{equation}
   G^{-1} (k) = i k_0 \gamma_0 + i v k_i \gamma_i  +  \Sigma (k),
   \end{equation}
   where the $d+1$-momentum $k =(k_0,\vec{k})$, we can write both contributions together as
   \begin{eqnarray}
   \Sigma(k) = i e_d ^2 \int \frac{d q}{ (2\pi)^{d+1} } \frac{q_\alpha
   +k_\alpha}{ |\vec{q}|^\epsilon (q+k)^2} \{ - \gamma_0 \gamma_\alpha
   \gamma_0 + \\ \nonumber 2 g ( \gamma_\mu \gamma_0 \gamma_\nu
   \gamma_\alpha \gamma_0 + \gamma_0 \gamma_\alpha \gamma_\nu \gamma_0
   \gamma_\mu) I_{\mu \nu} (q) \},
   \end{eqnarray}
   where $dq = dq_0 d^d \vec{q}$, and the integral
   \begin{equation}
   I_{\mu\nu}  (q) =  \int \frac{d p}{  (2\pi)^{d+1} } \frac{p_\mu (p+q)_\nu}{p^2 (p+q)^2  }.
   \end{equation}
   We have set the Fermi velocity in the last two equations to $v=1$
   for notational simplicity.  Since the integral over the momentum
   $q$ in the first term is only logarithmically divergent in the
   infrared, for the divergent part of the self-energy we need to
   include only the $q=0$ limit of $I_{\mu\nu} (q)$.  In any $d>1$ the
   integral $I_{\mu\nu}(q)$ is infrared convergent and may be easily
   computed to be
   \begin{equation}
   I_{\mu\nu} (q)= \frac{ \Lambda^\epsilon \delta_{\mu \nu} }{4\pi \epsilon} +O(q^\epsilon)
   \end{equation}
   when $\epsilon \ll 1$. On the other hand,
   \begin{equation}
   \delta_{\mu\nu} ( \gamma_\mu \gamma_0 \gamma_\nu \gamma_\alpha
   \gamma_0 + \gamma_0 \gamma_\alpha \gamma_\nu \gamma_0 \gamma_\mu)=
   2(1-d) \gamma_0 \gamma_\alpha \gamma_0,
   \end{equation}
   where we analytically continued the identity $\gamma_i \gamma_i =
   d$ to arbitrary spatial dimension $d$.  The self-energy therefore
   simplifies into
\begin{equation}
 \Sigma(k) = -i e_d ^2 \gamma_0 \gamma_\alpha \gamma_0  (  1+ \frac{g \Lambda^\epsilon}{\pi} )
 \int \frac{d q}{  (2\pi)^{d+1} } \frac{q_\alpha +k_\alpha}{ |\vec{q}|^\epsilon (q+k)^2},
 \end{equation}
 and becomes simply proportional to the leading contribution. The
 remaining integral may be performed in arbitrary dimension $d>1$, and
 it yields
 \begin{equation}
 \Sigma(k) = \frac{(1-d^{-1} ) e_d ^2 } { \Gamma(d/2) 2^d \pi^{d/2} }
 (1+ \frac{g \Lambda^\epsilon}{\pi} ) \ln( \frac{ \Lambda}{k} ) i k_i
 \gamma_i .
 \end{equation}
 In the limit $d\rightarrow 1$ the diverging part of the self-energy
 therefore becomes
 \begin{equation}
 \Sigma(k) = \frac{e^2}{\pi} (1+ \hat{g})\ln( \frac{ \Lambda}{k} ) i k_i \gamma_i.
 \end{equation}
Replacing the Gross-Neveu coupling $\hat{g}$ with its $O(\epsilon)$ critical value the last result may be
recast as in Eq. (10).

    \begin{figure}[t]
{\centering\resizebox*{50mm}{!}{\includegraphics{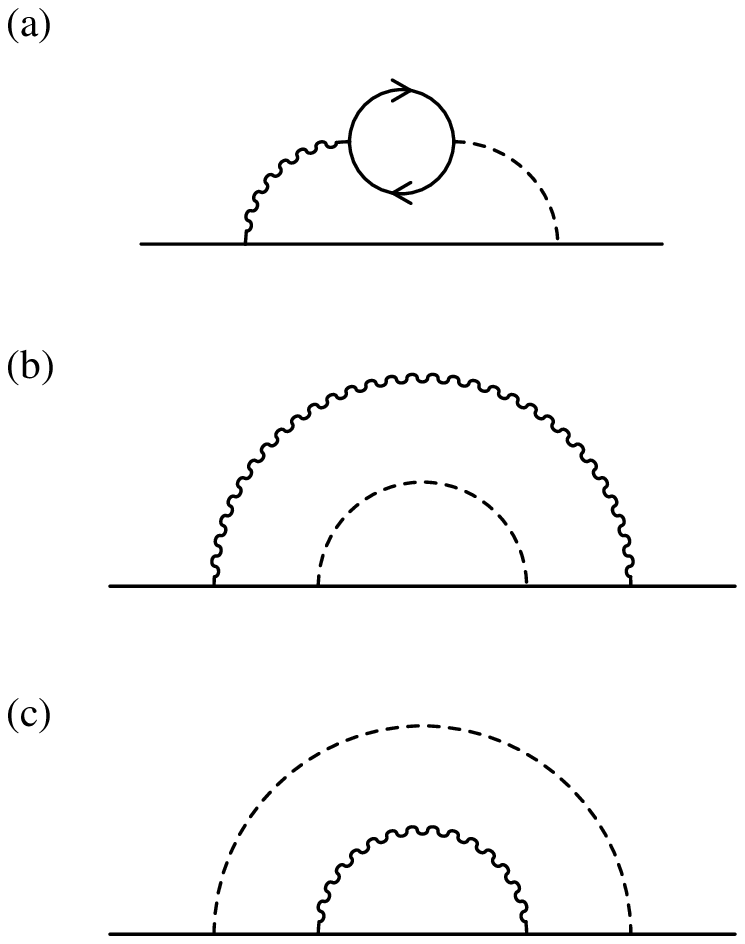}}
\par} \caption[] {The remaining three two-loop diagrams that do not contribute to the Fermi
velocity renormalization.}
\end{figure}

We turn now to the calculation of the beta-function of the short-range coupling in presence of the long-range
Coulomb interaction. Since the Coulomb coupling violates effective Lorentz invariance of the Gross-Neveu
Lagrangian, the engineering dimension of the short-range coupling constant receives an additional contribution
from a non-trivial dynamical exponent $z$, and it is equal to $\epsilon-(z-1)$. The beta-function of the
short-range coupling then becomes
\begin{equation}\label{GN-beta1}
\frac{d\hat{g}}{d\ln\Lambda}=(\epsilon-z+1){\hat g}-2(2N-1){\hat g}^2+c_d{\hat e}^2{\hat g}+O(\hat{g}^2{\hat
e}^2,\hat{g}^3),
\end{equation}
where the third term arises from the diagrams shown in Fig.\ 4, and the term proportional to ${\hat g}^2$
yields the usual beta-function for the Gross-Neveu coupling in the absence of other interactions
\cite{herbut1,herb-jur-roy}. In fact, only the diagram in Fig.\ 4a renormalizes the short-range coupling,
$g\rightarrow g+\delta g $, where
\begin{equation}
\delta g=-\frac{4\pi}{\epsilon}{\hat e}^2g\int\frac{dq}{(2\pi)^{d+1}}\frac{1}{(q+k)^2|{\vec q}|^\epsilon}.
\end{equation}
This integral may be performed in arbitrary dimension $d>1$, similarly to the one in Eq.\ (17), and in the
limit $d\rightarrow1$ its diverging part has the form
\begin{equation}
\delta g=-\frac{2}{\epsilon}{\hat e}^2 g \ln(\frac{\Lambda}{k}),
\end{equation}
yielding $c_d=-2/\epsilon$ in Eq.\ (\ref{GN-beta1}). The remaining diagram in Fig.\ 4b does not renormalize the
Gross-Neveu coupling, but generates a new short-range coupling of the form
$({\bar\Psi}\gamma_0{\vec\gamma}\Psi)^2$, which is irrelevant close to the critical point. An analogous
situation arises in the $(2+1)$-dimensional quantum electrodynamics with the additional Gross-Neveu interaction
\cite{kaveh-herbut}. The irrelevance of the generated short-range interaction also agrees with the emergent
Lorentz invariance in graphene \cite{herb-jur-roy}. Finally, identifying the right-hand side of Eq.\ (8) as
being precisely $(z-1) e_d ^2$ \cite{herbut1} yields the beta-function of the Gross-Neveu coupling as in Eq.\
(\ref{GN-beta}).

\begin{figure}[tb]
{\centering\resizebox*{50mm}{!}{\includegraphics{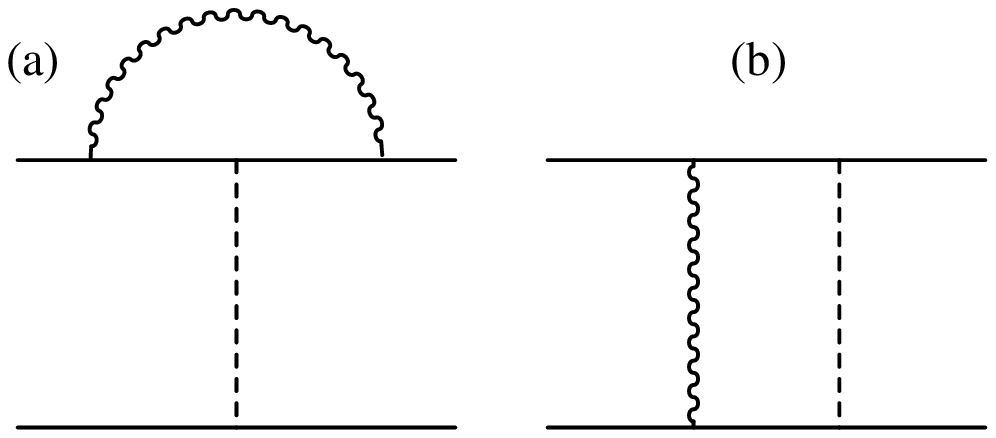}}
\par} \caption[] {Renormalization of the Gross-Neveu coupling due to
the long-range Coulomb interaction.}
\end{figure}

Remarkably, it is possible to obtain the explicit solution of the flow-equations in Eqs. (\ref{GN-beta}) and
(\ref{Coulomb-beta}). Since the long-range Coulomb interaction is marginally irrelevant in the infrared, it is
convenient to cast the two flow equations in the form
\begin{equation}\label{flow-diff}
e^4 d{g}/d{e}^2=\epsilon g -2(2N-1) g^2  -(2/\epsilon-1)g e^2,
\end{equation}
where we dropped the hats on the dimensionless couplings, and neglected the term $\sim\epsilon$ in Eq.\
(\ref{Coulomb-beta}). General solution of this differential equation reads
\begin{equation}\label{flow}
g(e^2)=g_c \frac{(\epsilon/e^2)^{(2/\epsilon-1)}{\rm e}^{-\epsilon/e^2}}{C+\Gamma(2/\epsilon,\epsilon/e^2)},
\end{equation}
with $C$ as the integration constant, $\Gamma(a,b)\equiv\int_b^\infty t^{a-1}{\rm e}^{-t}dt$. Depending on the
initial condition, Eq.\ (\ref{flow}) describes the infrared flow of the short-range coupling towards either
zero or infinity. The trajectory ending exactly at the critical point as the charge flows to zero,
$g(e^2\rightarrow 0)=g_c$, separates two regions with different asymptotic behavior of the flow in the
infrared, as in Fig.\ 1. The separatrix is obtained by setting $C=0$ in Eq.\ (\ref{flow}), so that the solution
near the Gross-Neveu critical point becomes
\begin{equation}
\frac{g(e^2)}{g_c}=1-\frac{2-\epsilon}{\epsilon^2}e^2+O(e^4).
\end{equation}
The separatrix therefore lies in the region $g<g_c$, and approaches the critical value of the Gross-Neveu
coupling as the charge flows to zero {\it linearly}.

To summarize, we have computed the beta-function governing the flow of the long-range tail of the Coulomb
interaction at the Gross-Neveu metal-insulator quantum critical point in an expansion near one spatial
dimension. The effect of the short-range Gross-Neveu interaction is to render the charge more (marginally)
irrelevant than at the Gaussian fixed point.  We discussed how the form of the flow diagram implies that the
(irrelevant) charge is nevertheless a possible tuning parameter for the metal-insulator transition in graphene.

This research was  supported in part by the National Science Foundation under Grant No. PHY05-51164, and by
NSERC of Canada.
 G.W.S.~acknowledges hospitality of the Galileo Galilei Institute
 and the Aspen Center for Physics.

\end{document}